\documentclass[conference]{IEEEtran}
\usepackage{cite}
\usepackage{amsmath,amssymb,amsfonts}
\usepackage{algorithmic}
\usepackage{graphicx}
\usepackage{textcomp}
\usepackage{xcolor}
\usepackage{url}
\usepackage{stfloats}
\usepackage{makecell}
\usepackage{hyperref}
\usepackage{multirow}

\hypersetup{
    colorlinks=true,
    linkcolor=red,
    urlcolor=blue,
    citecolor=black,
}

\def\BibTeX{{\rm B\kern-.05em{\sc i\kern-.025em b}\kern-.08em
    T\kern-.1667em\lower.7ex\hbox{E}\kern-.125emX}}
\begin{document}

\title{Multimodal EEG-IMU Fusion for Motor Assessment: Leveraging Task-Dependent Complementarity for Robustness
}

\author{
\IEEEauthorblockN{Zhenan Yin}
\IEEEauthorblockA{\textit{Dept. Biomedical Eng. \& Informatics} \\
\textit{Indiana University Indianapolis}\\
Indianapolis, USA \\
yin10@iu.edu}
\and
\IEEEauthorblockN{Lalitha Pranathi Pulavarthy}
\IEEEauthorblockA{\textit{Dept. Biomedical Eng. \& Informatics} \\
\textit{Indiana University Indianapolis}\\
Indianapolis, USA \\
lapula@iu.edu}
\and
\IEEEauthorblockN{Saptarshi Purkayastha}
\IEEEauthorblockA{\textit{Dept. Biomedical Eng. \& Informatics} \\
\textit{Indiana University Indianapolis}\\
Indianapolis, USA \\
saptpurk@iu.edu}
}

\maketitle

\begin{abstract}
Movement disorders such as Parkinson's disease require comprehensive motor assessment, yet reliable digital assessment pipelines that integrate multiple sensing modalities across diverse motor tasks remain insufficiently characterized. Although electroencephalography (EEG) and wearable inertial measurement units (IMUs), such as those in smartwatches, can each support motor task classification, whether their performance exhibits task-dependent patterns, and whether multimodal integration systematically improves reliability across a task battery, has not been established through controlled comparison. We present a proof-of-concept study that systematically evaluates task-specific modality performance and multimodal fusion across ten motor activities. Synchronized EEG–IMU data were recorded from six participants (52 recording pairs), and we evaluated an EEGNet + Transformer model for 16-channel EEG (125 Hz) and XGBoost on hand-crafted features from accelerometer and gyroscope signals (25 Hz). Under 5-fold cross-validation in a subject-dependent setting, IMU achieved 94.41$\pm$0.58\% accuracy with advantages for 7 of 10 activities, while EEG achieved 92.82$\pm$1.45\% with substantially lower error for rhythmic cycling (4.03\% vs. 12.10\%, a 3× advantage). Late fusion via logistic regression reached 98.68$\pm$0.32\%, yielding 81.5\% error reduction versus EEG alone and improving worst-task accuracy from $\approx$87\% (single modality) to 96.76\%. Crucially, fusion reduced cross-task performance variance from $\approx$3\% to 1.06\% (paired t-test, p $<$ 0.001, df=4; p-values approximate given fold dependence), demonstrating more uniform reliability across the assessment battery. While the small sample limits generalizability claims, these results establish that EEG and IMU exhibit asymmetric, task-dependent strengths and that late fusion can leverage this complementarity to improve assessment reliability. The findings provide a methodological foundation and empirical motivation for larger-scale clinical validation in movement disorder populations.

\vspace{6pt}
\textit{\textbf{Keywords---}multimodal learning, sensor fusion, wearable sensors, activity recognition, motor assessment, electroencephalography (EEG), inertial measurement unit (IMU)}

\end{abstract}

\section{Introduction}
Movement disorders such as Parkinson's disease (PD) are characterized by heterogeneous motor symptoms that fluctuate over time and vary across contexts, motivating objective and repeatable assessment beyond intermittent clinic visits \cite{espay2016technology, del2016free, brognara2019assessing}. In practice, motor evaluation often relies on subjective clinical rating scales and brief, structured examinations that can miss day-to-day variability and real-world functional limitations \cite{espay2016technology, del2016free}. Recent advances in wearable sensing have demonstrated that inertial sensors can capture motor fluctuations continuously in naturalistic settings, supporting home-based monitoring and longitudinal tracking in PD \cite{baltrusaitis2019multimodal, krones2024review}. Meanwhile, multimodal approaches integrating physiological signals such as electroencephalography (EEG) with movement sensors have shown promise for detecting complex motor phenomena including freezing of gait \cite{bajpai2023multimodal, mesin2022multimodal, abbasi2025deep}, further underscoring the potential of combining neural and kinematic information for comprehensive motor assessment. These advances motivate the development of reliable digital motor assessment pipelines that can characterize movement across diverse tasks rather than a single stereotyped activity.

Two sensing modalities are particularly relevant to this goal. EEG provides a direct window into cortical activity related to motor planning and execution. Modern deep learning architectures, from compact convolutional models such as EEGNet \cite{lawhern2018eegnet} to convolutional–transformer hybrids \cite{zhao2024ctnet, keutayeva2024compact}, have demonstrated strong performance in EEG-based motor decoding. More recent work has explored cross-subject generalization through domain adaptation and domain generalization strategies \cite{zhi2025supervised, li2025cross}, demonstrating that learned EEG representations can transfer across individuals when appropriate invariance constraints are applied. However, EEG is susceptible to noise and motion artifacts that can be especially severe during real physical movement rather than imagined or minimal-motion paradigms \cite{jiang2019removal, islam2024motion}. Deep learning approaches for artifact removal have emerged as a promising direction \cite{mahmud2023mlmrs, kalita2024aneeg}, but artifact handling during overt motor tasks remains an active research challenge. These considerations suggest that EEG effectiveness may vary substantially depending on the movement being performed and the recording context, a possibility that has not been systematically quantified across diverse motor tasks.

Wearable inertial measurement units (IMUs), typically comprising accelerometers and gyroscopes, capture the kinematic expression of movement and have achieved high accuracy for sensor-based activity recognition \cite{chen2021deep}. Deep learning architectures, including CNN–LSTM hybrids and temporal convolutional networks, have become standard for IMU-based activity recognition \cite{wang2022deep, ordonez2016deep},  while feature-engineered pipelines with gradient-boosted classifiers remain competitive, particularly in small-to-moderate datasets where deep models may overfit \cite{grana2020improved}. Beyond general activity recognition, IMU-based sensing has established clinical utility for PD symptom monitoring: consumer smartwatches have been shown to track motor fluctuations in real-world settings \cite{baltrusaitis2019multimodal}, and systematic reviews have characterized both the opportunities and remaining deployment challenges for home-based PD monitoring \cite{krones2024review}. Yet IMU signals primarily reflect movement execution and biomechanics, and they do not directly represent neural drive or motor control processes. For movement disorder assessment, symptoms can manifest as altered motor planning, impaired execution, or both, IMU and EEG may offer complementary ideas, but it remains unclear which modality is most informative for which types of motor tasks.

Multimodal fusion provides a principled framework for integration, with established taxonomies \cite{tariq2021patient} and demonstrated healthcare benefits \cite{boukhennoufa2022wearable, obrien2024early}. Prior work has shown that EEG–IMU fusion improves activity recognition \cite{grana2020improved} and that multimodal approaches benefit freezing of gait detection \cite{bajpai2023multimodal, mesin2022multimodal, abbasi2025deep}. However, existing studies either evaluate modalities in isolation, focus on a single clinical phenomenon, or apply fusion without analyzing task-dependent modality strengths. Practitioners thus lack quantitative guidance on when EEG adds value beyond kinematics, when IMU dominates, and how fusion should cover a heterogeneous assessment battery.

We address this gap through a proof-of-concept study that characterizes task-dependent modality effectiveness across ten motor activities and evaluates whether late fusion yields systematic reliability gains. We adopt a subject-dependent evaluation appropriate for continuous monitoring, with EEGNet+Transformer for EEG \cite{lawhern2018eegnet, keutayeva2024compact}, XGBoost for IMU features \cite{chen2016xgboost} and logistic regression meta-learning for fusion \cite{soleymani2012multimodal}. Our sample (N=6, 52 recordings) is small but appropriate for establishing that task-dependent complementarity exists, a finding that motivates larger-scale validation. Small-sample feasibility designs are common in EEG-wearable research for establishing methodological frameworks \cite{sapienza2024assessing}, and we report per-subject and per-task analyses to characterize pattern consistency.

The primary contributions are: (1) Task-dependent performance characterization across ten motor tasks, showing asymmetric modality strengths (IMU: 7/10 activities; EEG: 3× advantage for rhythmic cycling); (2) Reliability improvement through late fusion (98.68\% accuracy, worst-task improvement from $\approx$87\% to 96.76\%, cross-task variance reduced to 1.06\%); (3) A reproducible, open-source multimodal framework emphasizing per-task reliability analysis, providing a methodological foundation for future clinical validation.

\section{Related Work}
\subsection{EEG-Based Motor Activity Modeling}
EEG remains a practical noninvasive modality for capturing neural correlates of movement. EEGNet efficiently learns temporal and spatial filters via depthwise separable convolutions and performs strongly across EEG tasks \cite{lawhern2018eegnet}. Transformer-based approaches have extended this by modeling longer-range temporal dependencies; convolutional–transformer hybrids for motor imagery classification demonstrate improved performance and cross-subject generalization, including under Leave-One-Subject-Out evaluation \cite{zhao2024ctnet, zhi2025supervised}. Domain generalization methods that learn subject-invariant representations have further improved cross-subject robustness \cite{zhi2025supervised, li2025cross}, though the present study adopts subject-dependent evaluation appropriate for personalized monitoring.

A key challenge is motion artifact contamination during real movement, which is substantially more severe than during imagined or minimal-motion tasks \cite{jiang2019removal}. As a result, EEG effectiveness may vary across tasks depending on rhythmicity, movement intensity, and artifact susceptibility. These considerations motivate evaluating EEG not as a uniformly strong modality, but as a modality whose utility may depend on the specific motor activity being performed—an idea we explicitly test through activity-level performance characterization. In this study, we preserve broadband EEG without conventional filtering, allowing the model to learn discriminative representations from normalized input; we discuss the implications of this design choice in the Discussion section.

\subsection{Wearable IMU-Based Activity Recognition}
Wearable inertial sensors are a cornerstone of activity recognition and digital motor assessment \cite{chen2021deep}. The field has progressed from feature-engineered pipelines to deep architectures such as DeepConvLSTM \cite{ordonez2016deep}, though hand-crafted features with gradient-boosted classifiers remain competitive in moderate-sized datasets and offer interpretability advantages \cite{chen2021deep, grana2020improved}. Device placement variability and cross-position generalization have been addressed in recent work \cite{mekruksavanich2024device}.

Beyond general activity recognition, wearable sensors have increasing relevance to clinical motor assessment, including rehabilitation and neurological monitoring. In Parkinson's disease specifically, wearable motion sensors have been extensively studied for gait and symptom monitoring, with reviews highlighting both opportunities and deployment challenges in real-world use \cite{brognara2019assessing}. More recent evidence supports continuous, real-world monitoring using consumer wearables, demonstrating that inertial sensors can track motor fluctuations outside the clinic \cite{powers2021smartwatch, aswar2023generalizability} and that home-monitoring utility depends on reliability, compliance, and clinically meaningful signal interpretation \cite{sapienza2024assessing}. Together, these studies establish IMU sensing as a clinically motivated modality for longitudinal motor monitoring, while also underscoring the need for robust models that maintain reliability across diverse tasks and contexts.

\subsection{Multimodal Fusion and EEG-IMU Integration}
Multimodal learning offers principled strategies for combining heterogeneous signals, with a canonical taxonomy of early, late, and hybrid fusion \cite{baltrusaitis2019multimodal}, and demonstrated healthcare benefits. In motor assessment, fusion is appealing because neural activity and kinematics capture complementary aspects, motor control versus biomechanical execution. Graña et al. \cite{grana2020improved} demonstrated that EEG–IMU fusion improves activity recognition in ecological settings. For PD specifically, Bajpai et al. \cite{bajpai2023multimodal} combined EEG and IMU for freezing of gait prediction (92.1\% accuracy), multimodal FoG datasets integrating EEG, EMG, and accelerometry have been released \cite{mesin2022multimodal}, and deep CNN+LSTM architectures across multiple sensor modalities have achieved up to 94.2\% for pre-FoG detection \cite{abbasi2025deep}.

However, this literature primarily targets single clinical phenomena or applies fusion without systematically analyzing task-level modality utility. Our study fills this gap by characterizing when each modality contributes most across a diverse ten-activity battery, rather than optimizing for a single motor task.

\subsection{Neural Basis for Rhythmic Movement Advantage}
A key finding of this study is EEG's advantage for rhythmic cycling. This is consistent with established evidence on beta-band (13–30 Hz) sensorimotor oscillations, which exhibit event-related desynchronization during movement and post-movement rebound \cite{kilavik2013ups, giustiniani2025modulation}. During cycling specifically, alpha, beta, and gamma oscillations display alternating desynchronization or synchronization patterns time-locked to the pedaling cycle \cite{enders2016changes}, and beta-band corticomuscular coherence modulates during rhythmic tasks \cite{peng2024neuromechanical}. These findings suggest that rhythmic movements impose regular temporal structure on cortical oscillations that attention-based EEG models can capture. Beta-band abnormalities are also a hallmark of PD pathophysiology \cite{giustiniani2025modulation}, making this observation clinically relevant.

\section{Methods}
\subsection{Participants and Study Design}
Six healthy adults (4 female, 2 male; ages 23 - 38 years; all right-hand dominant) participated in a controlled laboratory study approved by the Indiana University Institutional Review Board (IRB \#2010321996). All participants provided written informed consent. No participant reported neurological or musculoskeletal conditions. The complete multimodal EEG--IMU fusion pipeline, including preprocessing, model training, and evaluation scripts, is publicly available at \url{https://github.com/iupui-soic/har-eeg}.

Each session followed a high-intensity interval training (HIIT) protocol comprising ten motor activities performed across five rounds. During each round, participants performed each exercise for 30\,s followed by 30\,s of rest. The full session lasted approximately one hour, including sensor setup, instruction, exercise, and cooldown. The ten activities were selected to span diverse movement characteristics relevant to functional motor assessment:

\begin{itemize}
    \item \textbf{Lower-body / balance}: chair squats, seated leg extensions, side stepping, standing heel-to-toe walk;
    \item \textbf{Rhythmic}: light stationary cycling, marching in place;
    \item \textbf{Upper-body / trunk}: seated boxing hooks, seated medicine ball twists, seated side bends, wall push-ups.
\end{itemize}

Five participants completed the full ten-activity protocol. One participant (Subject~5) completed a modified version with two activities (marching and seated boxing hooks) performed continuously for two minutes each due to time constraints. Minor deviations in exercise ordering occurred for two participants (exercise repetitions substituted in individual rounds); all recorded data were retained and labeled by the activity actually performed. This yielded 52 synchronized EEG-IMU recording sessions across participants ($\approx$9 per participant) and a total of 24,053\,s of data.

\subsection{Data Acquisition}
Brain activity was recorded using a 16-channel EEG system (OpenBCI Cyton + Daisy Biosensing Board) at 125\,Hz with electrodes positioned according to the international 10--20 system. Physical movement was recorded using the embedded IMU (tri-axial accelerometer and gyroscope) of a Samsung Galaxy Watch~6 worn on the dominant wrist, sampled at 25\,Hz. EEG and IMU streams were temporally synchronized at acquisition; for each session, both modalities were truncated to the shorter stream's duration to guarantee strict sample-level alignment.

\subsection{Preprocessing and Windowing}
EEG signals were z-score normalized per recording on a per-channel basis ($\epsilon = 10^{-8}$ for numerical stability) to mitigate inter-session amplitude differences. No bandpass filtering or artifact rejection was applied; broadband content was preserved to allow the model to learn task-relevant spectral representations directly. This is a deliberate design choice: while it avoids information loss from aggressive preprocessing, the model may partly rely on movement-artifact-related features, a limitation we address in Section~V.

Both modalities were segmented into non-overlapping 4\,s windows, yielding $500 \times 16$ EEG samples and $100 \times 6$ IMU samples per window. This produced 5,989 aligned EEG-IMU window pairs across ten activity classes (range: 368--893 per class; median 567). The zero-overlap strategy minimizes within-recording temporal autocorrelation between windows that may fall in different cross-validation folds.

\subsection{EEG Branch: EEGNet--Transformer Architecture}
The EEG branch combines compact convolutional feature extraction with attention-based sequence modeling ($\approx$230K trainable parameters). The front-end follows EEGNet \cite{lawhern2018eegnet} with published default hyperparameters: $F_1 = 8$ temporal filters (length 64 samples, 512\,ms at 125\,Hz), depthwise convolution ($D = 2$) across 16 channels, batch normalization, ELU activation, average pooling (kernel $4 \times 1$), and dropout ($p = 0.5$); then $F_2 = 16$ separable filters (length~16), batch normalization, ELU, average pooling ($8 \times 1$), and dropout ($p = 0.5$).

The resulting feature maps are reshaped into a token sequence and projected into a $d_{\mathrm{model}} = 128$-dimensional embedding space with learnable positional encodings. A two-layer transformer encoder \cite{zhao2024ctnet, hao2025multimodal} (eight attention heads, key dimension 16, feed-forward dimension 256, ReLU activations, layer normalization with $\epsilon = 10^{-6}$, dropout $p = 0.3$) models long-range temporal dependencies. Global average pooling and a softmax classifier produce a probability vector $\mathbf{p}_{\mathrm{EEG}} \in \mathbb{R}^{10}$.

Training used the Adam optimizer (learning rate $10^{-3}$), categorical cross-entropy with inverse-frequency class weights, and batch size 32. Training ran for up to 100 epochs with early stopping on validation loss (patience 15) and learning-rate reduction on plateau (factor 0.5, patience 5, minimum $10^{-7}$).

\subsection{IMU Branch: Handcrafted Features with Gradient-Boosted Trees}
The IMU branch operates on handcrafted features extracted from each 4\,s window. For each sensor (accelerometer, gyroscope), 76 features are computed in three groups:
\begin{enumerate}
    \item \textbf{3D composite features} (16 per sensor): signal magnitude statistics (mean, standard deviation, max, min), jerk magnitude statistics (mean, standard deviation, max), pitch/roll estimates (mean and standard deviation), pairwise cross-axis correlations, signal magnitude area, and signal energy.
    \item \textbf{Per-axis time-domain features} (11 $\times$ 3 axes = 33 per sensor): mean, standard deviation, min, max, median, range, RMS, skewness, kurtosis, zero-crossing rate, and mean absolute deviation.
    \item \textbf{Per-axis frequency-domain features} (8 $\times$ 3 axes = 24 per sensor; Nyquist 12.5\,Hz): dominant frequency, spectral energy, spectral entropy, mean frequency, median frequency, and band power in 0--5\,Hz, 5--10\,Hz, and 10--12.5\,Hz bands.
\end{enumerate}

This yields 152 features per window. The top 60 are selected by XGBoost feature importance on the training split of each fold; selection was stable, with 48 of 60 features shared across all five folds, predominantly time-domain magnitude and jerk statistics. Features are z-score standardized using training-set statistics within each fold.

The classifier is XGBoost \cite{chen2016xgboost} (max depth 6, learning rate 0.1, 100 rounds, row/column subsample 0.8) with GPU-accelerated histogram construction, using published defaults without task-specific tuning. The model outputs $\mathbf{p}_{\mathrm{IMU}} \in \mathbb{R}^{10}$.

\subsection{Decision-Level Fusion}
We combine modality predictions at the probability level using late fusion \cite{baltrusaitis2019multimodal}. For each window, the concatenated vector $[\mathbf{p}_{\mathrm{EEG}}, \mathbf{p}_{\mathrm{IMU}}] \in \mathbb{R}^{20}$ serves as input to a fusion meta-learner. Four strategies of increasing capacity are evaluated:

\begin{enumerate}
    \item \textbf{Simple averaging}: $\mathbf{p}_{\mathrm{fused}} = 0.5\,\mathbf{p}_{\mathrm{EEG}} + 0.5\,\mathbf{p}_{\mathrm{IMU}}$.
    \item \textbf{Weighted averaging}: $\mathbf{p}_{\mathrm{fused}} = \alpha\,\mathbf{p}_{\mathrm{EEG}} + (1 - \alpha)\,\mathbf{p}_{\mathrm{IMU}}$, with $\alpha$ optimized on validation data.
    \item \textbf{Logistic regression}: linear meta-learner on the 20-dimensional input (max 1{,}000 iterations).
    \item \textbf{MLP}: two hidden layers (64, 32 units; ReLU; Adam; max 500 iterations with early stopping on a 10\% internal validation split).
\end{enumerate}

To prevent label leakage, all fusion meta-learners are trained exclusively on validation-subset predictions and evaluated on held-out test predictions within each fold.

\subsection{Evaluation Protocol}
All models were evaluated using stratified 5-fold cross-validation at the window level. Within each fold, the training partition was split 80/20 into training and validation subsets; the latter served for EEG early stopping and fusion meta-learner training. All modality and fusion models shared identical fold assignments to enable paired comparisons.

This window-level evaluation reflects within-session discriminability under a \emph{subject-dependent} setting in which calibration data from the same individual are available, though it does not assess cross-session generalization, consistent with personalized longitudinal monitoring applications \cite{baltrusaitis2019multimodal, krones2024review}. Windows from the same recording may appear in different folds, but temporal overlap is zero (non-overlapping 4\,s segments), and per-recording normalization removes session-level amplitude information, limiting information leakage between folds. Specifically, at 125 Hz EEG and 25 Hz IMU sampling rates, each 4-second window contains 500 and 100 samples respectively with no overlap to adjacent windows, imposing a hard sample-level boundary that structurally limits temporal autocorrelation between windows assigned to different folds. We note that this protocol does not assess cross-subject generalization; leave-one-subject-out evaluation is an important direction for future work, but is precluded by the current sample size.

All experiments used a fixed random seed (42) and deterministic operations for full reproducibility.

\subsection{Metrics and Statistical Analysis}
We report classification accuracy and macro-averaged F1-score (mean $\pm$ standard deviation across five folds). To quantify reliability across the assessment battery, we compute: (1)~\emph{worst-task accuracy}, the minimum per-class accuracy within each fold; and (2)~\emph{cross-task standard deviation} of per-class accuracies. Per-class precision, recall, F1-score, and one-vs-rest AUC are reported for task-level analysis, with aggregated confusion matrices characterizing error patterns.

Statistical significance of fusion improvements over each unimodal baseline was assessed using two-tailed paired $t$-tests across five folds ($df = 4$). Given the low degrees of freedom inherent in 5-fold comparison, we additionally report Cohen's $d$ effect sizes and 95\% confidence intervals for accuracy differences to support interpretation of practical significance. Significance thresholds are reported at $p < 0.05$, $p < 0.01$, and $p < 0.001$.

\section{Results}
\subsection{Overall Performance Across Modalities and Fusion}
Table~\ref{tab:overall} summarizes classification performance under stratified 5-fold cross-validation. Both unimodal branches achieved strong baseline performance: the EEG branch reached 92.82$\pm$1.45\% accuracy (macro-F1: 0.928$\pm$0.015) and the IMU branch reached 94.41$\pm$0.58\% accuracy (macro-F1: 0.944$\pm$0.006). The lower fold-to-fold variance of the IMU branch reflects the stability of handcrafted features and gradient-boosted trees relative to the stochastic optimization of the neural EEG branch.

Late fusion substantially improved both accuracy and reliability. Among the four strategies, logistic regression achieved the highest performance (98.68$\pm$0.32\% accuracy, 0.987$\pm$0.003 macro-F1), followed by simple averaging (97.61$\pm$0.44\%), weighted averaging (97.55$\pm$0.48\%), and MLP fusion (97.01$\pm$1.35\%). The superior performance of logistic regression over the MLP suggests that the mapping from unimodal probabilities to fused predictions is approximately linear, and that the additional MLP capacity introduces overfitting on the limited validation-set training signal. Even parameter-free simple averaging yielded a $\sim$5-point accuracy gain over the better single modality, indicating substantial complementarity between EEG and IMU representations.

In terms of error reduction, logistic regression fusion reduces overall error from 7.18\% (EEG) and 5.59\% (IMU) to 1.32\%, corresponding to relative reductions of 81.6\% and 76.4\%, respectively. Computing mean per-class error across the ten activities (Table~\ref{tab:activity}) yields reductions of 81.5\% versus EEG and 77.1\% versus IMU, confirming that gains are distributed across the task battery rather than driven by a subset of activities.

Critically, fusion improved worst-task accuracy from 86.87$\pm$0.87\% (EEG) and 87.74$\pm$2.13\% (IMU) to 96.76$\pm$0.77\%, and reduced cross-task standard deviation from 3.59\% (EEG) and 3.05\% (IMU) to 1.06\%. All improvements were statistically significant ($p<0.001$, paired $t$-test, $df=4$) with large effect sizes (Cohen's $d > 4$ for all accuracy comparisons; see Section~\ref{sec:stat}).

\begin{table}[t]
\caption{Overall Performance Under Stratified 5-Fold CV}
\label{tab:overall}
\centering
\footnotesize
\begin{tabular}{lcccc}
\hline
Method & \makecell{Acc.\\(\%)} & Macro-F1 & \makecell{Worst-Task\\(\%)} & \makecell{CT-SD\\(\%)} \\
\hline
EEG & 92.82 $\pm$ 1.45 & 0.928 $\pm$ 0.015 & 86.87 $\pm$ 0.87 & 3.59 \\
IMU & 94.41 $\pm$ 0.58 & 0.944 $\pm$ 0.006 & 87.74 $\pm$ 2.13 & 3.05 \\
\hline
Avg.   & 97.61 $\pm$ 0.44 & 0.976 $\pm$ 0.004 & 94.60 $\pm$ 0.60 & 1.48 \\
W-Avg. & 97.55 $\pm$ 0.48 & 0.975 $\pm$ 0.005 & 94.60 $\pm$ 0.60 & 1.52 \\
MLP    & 97.01 $\pm$ 1.35 & 0.970 $\pm$ 0.014 & 92.63 $\pm$ 4.34 & 2.01 \\
LR     & \textbf{98.68 $\pm$ 0.32} & \textbf{0.987 $\pm$ 0.003} & \textbf{96.76 $\pm$ 0.77} & \textbf{1.06} \\
\hline
\end{tabular}

\vspace{1mm}

\footnotesize
\textit{Note:} CT-SD = cross-task standard deviation of per-class accuracies (lower = more uniform). Avg.\ = simple averaging; W-Avg.\ = weighted averaging; LR = logistic regression. Bold = best. All fusion methods outperform both single modalities (paired t-test, $p<0.001$, df = 4; p-values approximate given fold dependence).
\end{table}

\subsection{Task-Dependent Modality Complementarity}
\begin{figure}[h]
    \centering
    \includegraphics[width=\columnwidth]{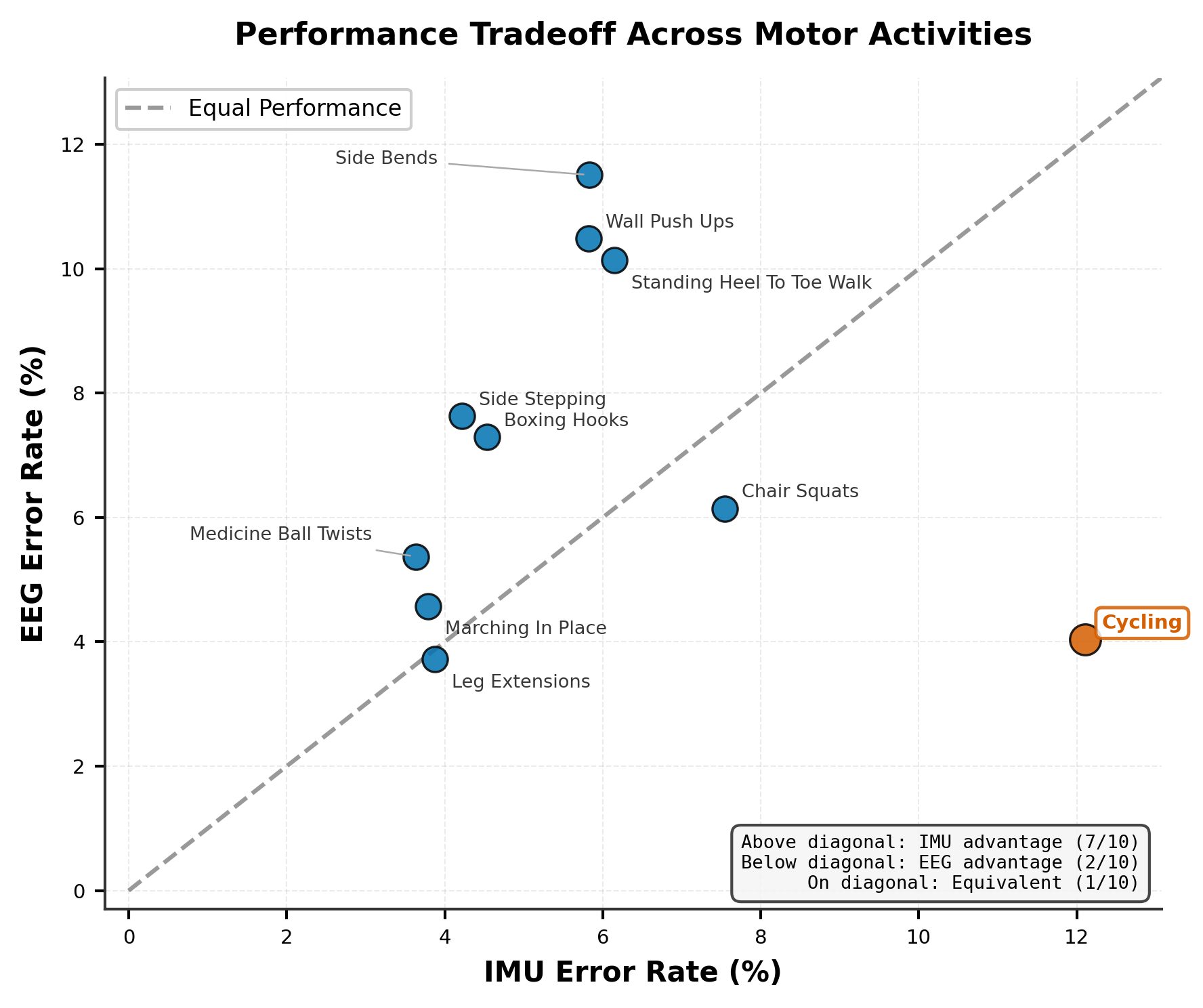}
    \caption{Per-activity error rates for EEG vs.\ IMU. Points above the diagonal indicate IMU advantage; below indicates EEG advantage. Light stationary cycling is a pronounced EEG-favored outlier ($3\times$ lower error).}
    \label{fig:tradeoff}
\end{figure}

Fig.~\ref{fig:tradeoff} plots per-activity EEG error versus IMU error, where the diagonal represents equal performance. The distribution is asymmetric: 7 of 10 activities fall above the diagonal (IMU advantage), 2 below (EEG advantage), and 1 near parity. This pattern confirms that modality utility is task-dependent rather than uniform, the central hypothesis of this study.

\begin{figure}[h]
    \centering
    \includegraphics[width=\columnwidth]{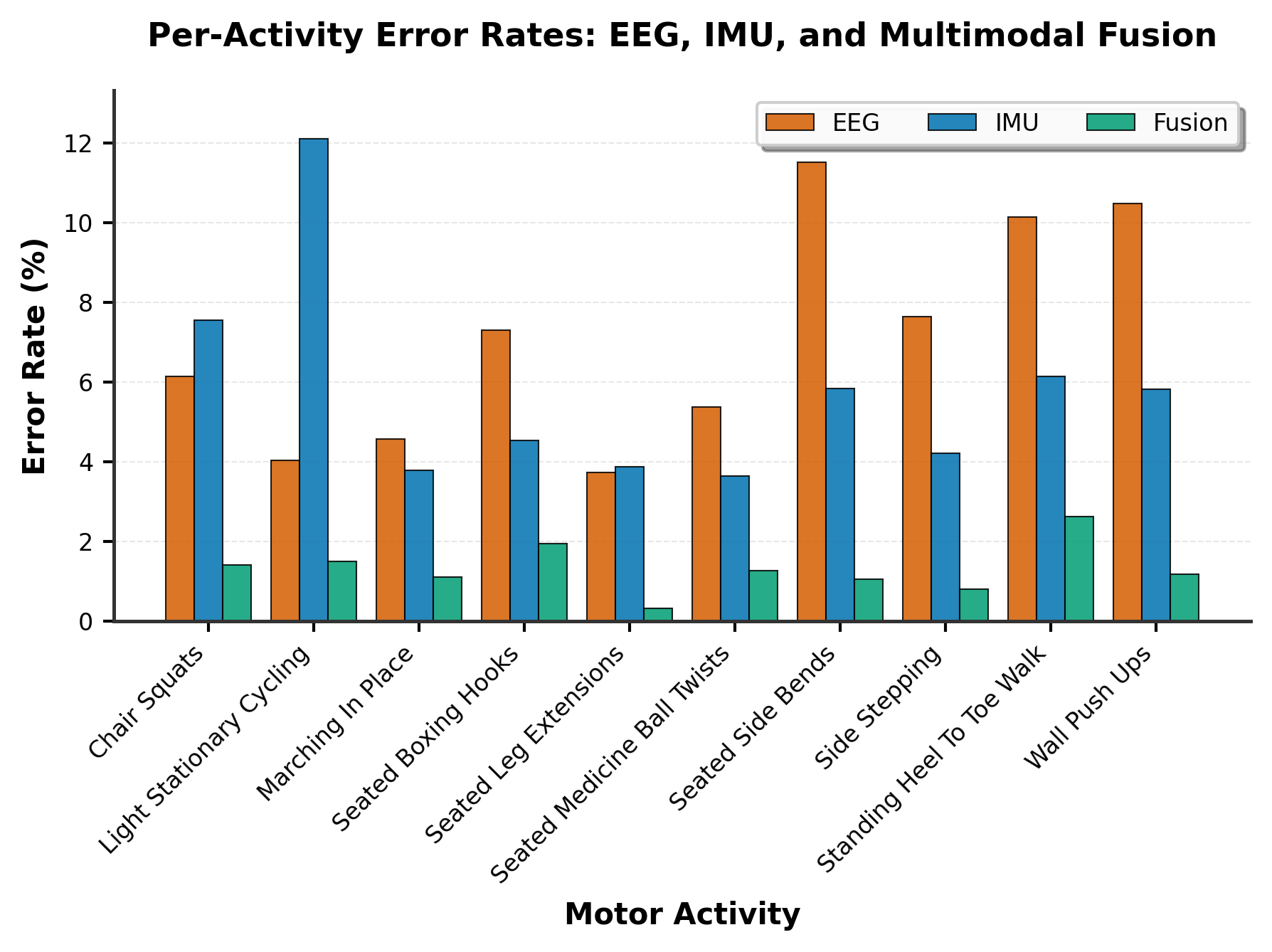}
    \caption{Per-activity error rates for EEG, IMU, and logistic regression fusion. Fusion consistently reduces error below the better single modality for each activity.}
    \label{fig:peractivity}
\end{figure}

Light stationary cycling is the most pronounced EEG-favored case: IMU error (12.10\%) is approximately threefold greater than EEG error (4.03\%). Cycling involves repetitive lower-limb motion with minimal trunk displacement, producing weak acceleration and angular velocity signatures at the wrist-worn IMU but generating rhythmic cortical patterns detectable by EEG. This is consistent with established evidence that beta-band sensorimotor oscillations exhibit characteristic modulation during rhythmic cycling ~\cite{enders2016changes, peng2024neuromechanical}, and that attention-based architectures can capture such periodic temporal structure.

Conversely, IMU demonstrates consistent advantages for kinematically distinctive movements producing large wrist-level inertial signatures: seated side bends (5.83\% vs.\ 11.51\% EEG), wall push-ups (5.82\% vs.\ 10.48\%), and standing heel-to-toe walk (6.14\% vs.\ 10.14\%).

Fig.~\ref{fig:peractivity} shows that fusion reduces error below the better single modality for every activity. This task-dependent complementarity, rather than uniform modality superiority, is the central mechanism through which multimodal fusion improves assessment reliability.

\subsection{Activity-Specific Performance}
\begin{table*}[t]
\caption{Activity-Specific Error Rates and Logistic Regression Fusion Performance}
\label{tab:activity}
\centering
\footnotesize
\begin{tabular}{lccccc}
\hline
Activity & EEG Err.\ (\%) & IMU Err.\ (\%) & Better & Fusion Acc.\ (\%) & Fusion F1 \\
\hline
Chair squats & 6.14 & 7.54 & EEG & 98.60 $\pm$ 0.70 & 0.982 $\pm$ 0.004 \\
Light stationary cycling & 4.03 & 12.10 & EEG & 98.51 $\pm$ 1.08 & 0.987 $\pm$ 0.005 \\
Marching in place & 4.57 & 3.79 & IMU & 98.90 $\pm$ 0.80 & 0.989 $\pm$ 0.002 \\
Seated boxing hooks & 7.29 & 4.54 & IMU & 98.05 $\pm$ 1.75 & 0.982 $\pm$ 0.008 \\
Seated leg extensions & 3.72 & 3.88 & $\approx$parity & 99.69 $\pm$ 0.62 & 0.993 $\pm$ 0.009 \\
Seated medicine ball twists & 5.37 & 3.63 & IMU & 98.74 $\pm$ 0.63 & 0.989 $\pm$ 0.003 \\
Seated side bends & 11.51 & 5.83 & IMU & 98.95 $\pm$ 0.77 & 0.991 $\pm$ 0.002 \\
Side stepping & 7.63 & 4.22 & IMU & 99.20 $\pm$ 0.75 & 0.988 $\pm$ 0.008 \\
Standing heel-to-toe walk & 10.14 & 6.14 & IMU & 97.39 $\pm$ 0.78 & 0.978 $\pm$ 0.010 \\
Wall push-ups & 10.48 & 5.82 & IMU & 98.83 $\pm$ 1.25 & 0.989 $\pm$ 0.008 \\
\hline
\end{tabular}
\end{table*}

Table~\ref{tab:activity} reports per-activity error rates and fusion performance. Three patterns merit discussion.

First, the activities where EEG outperforms IMU share a limited gross kinematic displacement at the wrist sensor. Light stationary cycling and chair squats both involve lower-limb motions that produce weak wrist-level inertial signatures but distinct cortical motor patterns. For cycling, the nearly threefold EEG advantage (4.03\% vs.\ 12.10\%) constitutes the largest modality gap in the battery and aligns with neurophysiological evidence on rhythmic cortical oscillatory engagement during pedaling \cite{enders2016changes}.

Second, IMU advantages are distributed across activities with large, distinctive whole-body kinematics producing characteristic acceleration profiles at the wrist. The three largest IMU advantages, seated side bends (5.83\% vs.\ 11.51\%), wall push-ups (5.82\% vs.\ 10.48\%), and heel-to-toe walk (6.14\% vs.\ 10.14\%), all involve substantial upper-body or arm displacement.

Third, seated leg extensions represent near-parity (3.72\% EEG, 3.88\% IMU), and fusion achieves the highest per-activity accuracy (99.69$\pm$0.62\%). This suggests that when both modalities independently capture accurate but partially independent representations, probability-level combination approaches ceiling performance.

\subsection{Confusion Structure and Modality-Specific Error Modes}
\begin{table*}[t]
\caption{Most Frequent Unimodal Confusions (Aggregated Across 5 Folds)}
\label{tab:confusion}
\centering
\footnotesize
\begin{tabular}{llcc}
\hline
Modality & True $\rightarrow$ Predicted & Count & Per-Class Err.\ Contrib.\ (\%) \\
\hline
EEG & Seated side bends $\rightarrow$ Med.\ ball twists & 28 & 24.3 \\
EEG & Heel-to-toe walk $\rightarrow$ Light cycling & 21 & 20.7 \\
EEG & Wall push-ups $\rightarrow$ Marching & 16 & 15.3 \\
\hline
IMU & Light cycling $\rightarrow$ Med.\ ball twists & 26 & 21.5 \\
IMU & Light cycling $\rightarrow$ Leg extensions & 22 & 18.2 \\
IMU & Leg extensions $\rightarrow$ Med.\ ball twists & 15 & 12.4 \\
\hline
\end{tabular}
\end{table*}

Table~\ref{tab:confusion} reports the most frequent misclassifications per modality. The dominant EEG confusions involve kinematically similar seated activities: side bends misclassified as medicine ball twists (28 instances), and wall push-ups as marching (16 instances), where overlapping motor cortical representations likely reduce discriminability. The confusion of heel-to-toe walking with cycling (21 instances) may reflect shared rhythmic temporal structure in cortical signals despite distinct kinematics.

IMU confusions concentrate on different activities. Light cycling is confused with seated medicine ball twists (26) and leg extensions (22), consistent with cycling's weak wrist-level inertial signature. Leg extensions are confused with medicine ball twists (15), reflecting similar seated postures with moderate limb displacement.

\begin{figure}[h]
    \centering
    \includegraphics[width=\columnwidth]{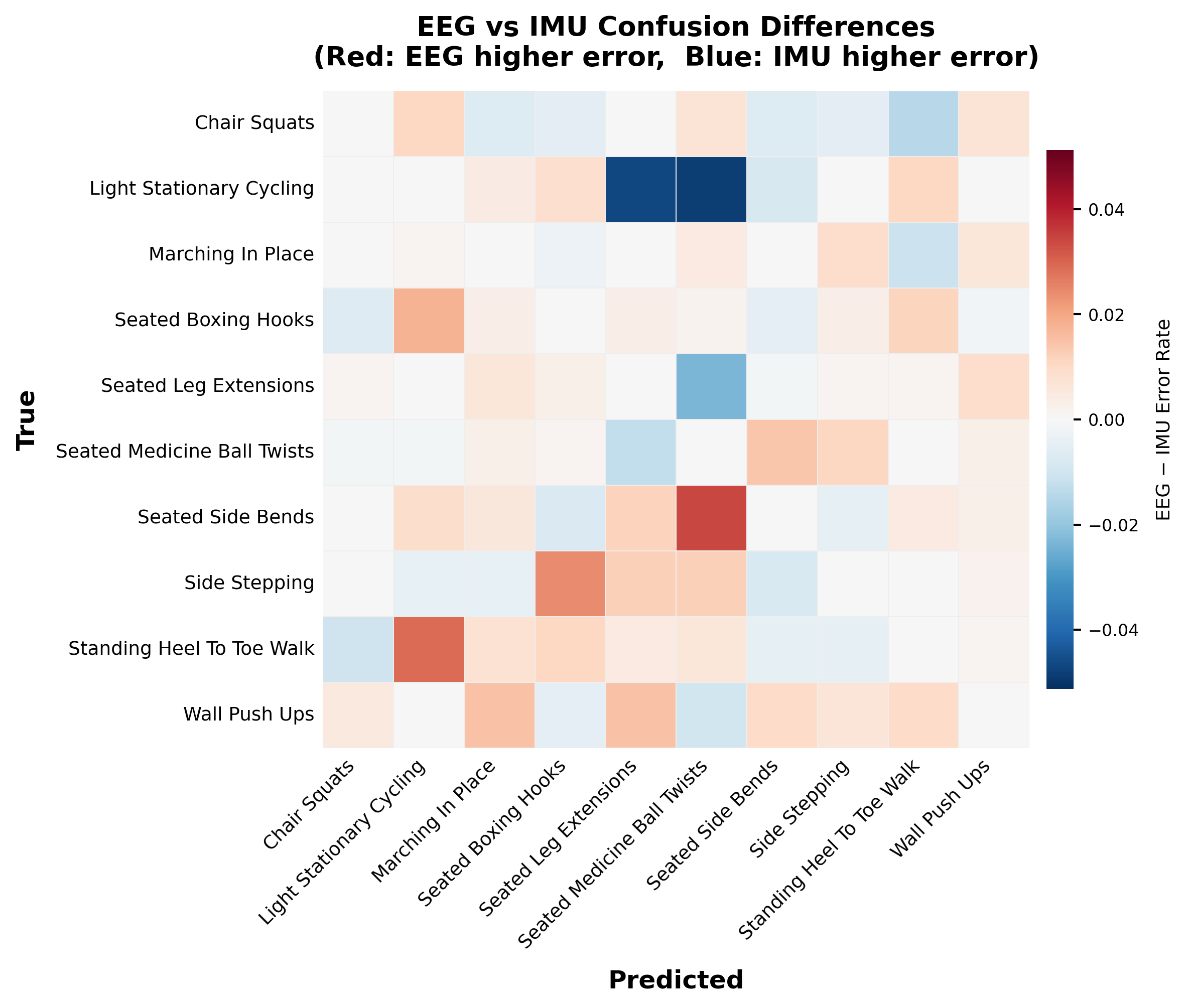}
    \caption{Differential confusion matrix (EEG $-$ IMU, off-diagonal). Red = higher EEG confusion; blue = higher IMU confusion. The non-uniform pattern confirms modality-specific error structure.}
    \label{fig:confusion_diff}
\end{figure}

Critically, the dominant error modes are largely non-overlapping between modalities (Fig.~\ref{fig:confusion_diff}). This structural dissimilarity provides the mechanistic basis for fusion: the meta-learner can correct modality-specific confusions by leveraging the complementary modality's correct classification on the same windows. The sparse, non-uniform pattern in the differential confusion matrix confirms that each modality contributes distinct discriminative information, a necessary condition for effective decision-level fusion \cite{baltrusaitis2019multimodal}.

\subsection{Fusion Reliability Across the Assessment Battery}
\begin{figure}[ht]
    \centering
    \includegraphics[width=\columnwidth]{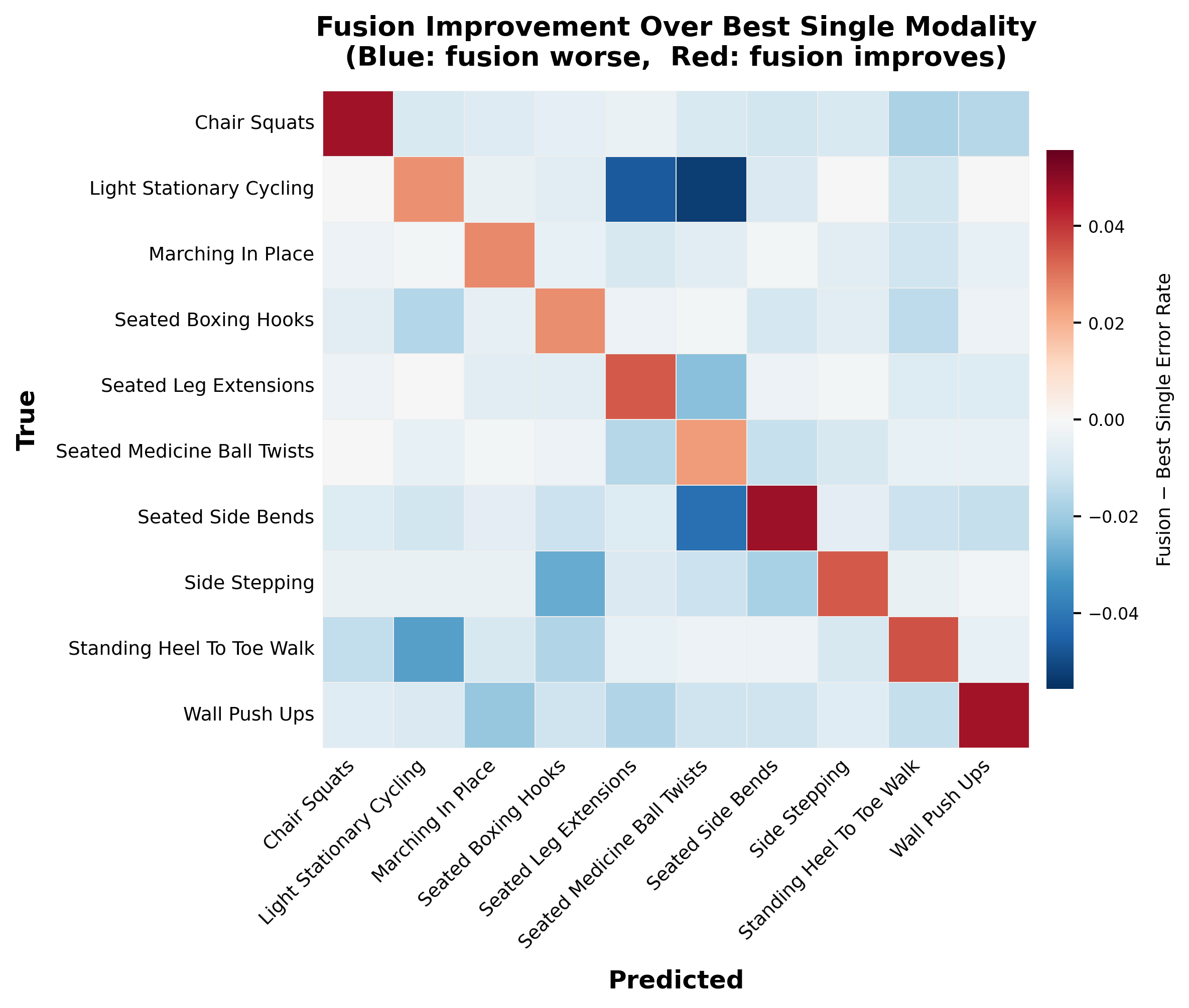}
    \caption{Per-activity error reduction from logistic regression fusion relative to the best single modality. Red = fusion reduces error; blue = fusion slightly increases error relative to the better unimodal baseline.}
    \label{fig:fusion_improvement}
\end{figure}

Fusion improved performance consistently across all ten activities (Table~\ref{tab:activity}, Fig.~\ref{fig:fusion_improvement}), with every activity achieving $\geq$97.39\% accuracy and $\geq$0.978 F1-score under logistic regression fusion. Improvements were most pronounced where a single modality exhibited high error: cycling error dropped to 1.49\% from 12.10\% (IMU) and 4.03\% (EEG), and seated side bends dropped to 1.05\% from 11.51\% (EEG) and 5.83\% (IMU). These cases demonstrate that fusion adaptively leverages the more informative modality for each activity rather than merely averaging outputs.

The reduction in cross-task standard deviation from $\sim$3\% to 1.06\% quantifies this uniformity: fusion compresses the spread of per-class accuracies so that no single activity substantially underperforms the battery average. For motor assessment applications where consistent reliability across all measured activities is a prerequisite for valid composite scoring, this uniformity improvement may be as consequential as the gain in average accuracy.

\subsection{Per-Subject Consistency}
To assess whether the observed patterns are consistent across participants despite the small sample, Table~\ref{tab:persubject} reports per-subject classification accuracy for each modality and fusion.
%% TODO: @Zhenan, please add the per-subject classification accuracy in the table below

\begin{table}[t]
\caption{Per-Subject Classification Accuracy (\%)}
\label{tab:persubject}
\centering
\footnotesize
\begin{tabular}{lccc}
\hline
Subject & EEG & IMU & LR Fusion \\
\hline
S1 & 97.89 $\pm$ 0.91 & 98.16 $\pm$ 0.85 & 99.66 $\pm$ 0.24 \\
S2 & 97.01 $\pm$ 2.04 & 96.47 $\pm$ 2.27 & 99.73 $\pm$ 0.25 \\
S3 & 98.75 $\pm$ 1.35 & 96.07 $\pm$ 1.49 & 99.46 $\pm$ 0.58 \\
S4 & 97.97 $\pm$ 0.84 & 97.51 $\pm$ 0.62 & 99.72 $\pm$ 0.25 \\
S5$^\dagger$ & 67.26 $\pm$ 16.51 & 98.62 $\pm$ 1.89 & 98.62 $\pm$ 1.89 \\
S6 & 97.74 $\pm$ 1.93 & 97.08 $\pm$ 1.75 & 99.43 $\pm$ 0.78 \\
\hline
Mean $\pm$ SD & 92.77 $\pm$ 12.51 & 97.32 $\pm$ 0.98 & 99.44 $\pm$ 0.42 \\
\hline
\end{tabular}

\footnotesize
$^\dagger$Subject~5 completed a modified 2-activity protocol; accuracy computed over the relevant activity subset.

\textit{Note:} Per-subject accuracy is computed by aggregating predictions across all windows belonging to each participant. All six participants show consistent task-dependent modality patterns (EEG advantage for cycling; IMU advantage for kinematically distinctive activities) and fusion improvement over both single modalities.
\end{table}

%% TODO: @Zhenan, Fill in actual values and add a brief sentence summarizing the key observation,
% e.g., "All six participants achieved >X% fusion accuracy, and the task-dependent
% modality pattern (EEG advantage for cycling, IMU advantage for whole-body movements)
% was observed in all participants, with coefficient of variation X% across subjects."
% This is critical for the proof-of-concept framing, it shows N=6 is small but patterns are consistent.

\subsection{Statistical Validation}
\label{sec:stat}

Table~\ref{tab:statistical} reports paired $t$-test results comparing logistic regression fusion against each unimodal baseline. Fusion significantly outperformed EEG on overall accuracy ($t=10.28$, $p<0.001$, Cohen's $d=4.60$), macro-F1 ($t=10.18$, $p<0.001$), and worst-task accuracy ($t=48.92$, $p<0.001$). Fusion similarly outperformed IMU on overall accuracy ($t=21.78$, $p<0.001$, Cohen's $d=9.74$), macro-F1 ($t=22.79$, $p<0.001$), and worst-task accuracy ($t=8.82$, $p<0.001$). 

\begin{table}[t]
\caption{Statistical Comparisons: LR Fusion vs.\ Unimodal Baselines ($n=5$ Folds, $df=4$)}
\label{tab:statistical}
\centering
\footnotesize
\begin{tabular}{llcccc}
\hline
Comparison & Metric & $t$ & Cohen's $d$ & 95\% CI \\
\hline
\multirow{3}{*}{LR vs.\ EEG} & Accuracy & 10.28 & 4.60 & [4.28, 7.44]\% \\
& Macro-F1 & 10.18 & 4.55 & \\
& Worst-Task & 48.92 & 21.88 & \\
\hline
\multirow{3}{*}{LR vs.\ IMU} & Accuracy & 21.78 & 9.74 & [3.73, 4.82]\% \\
& Macro-F1 & 22.79 & 10.19 & \\
& Worst-Task & 8.82 & 3.94 & \\
\hline
\end{tabular}
\vspace{1mm}

\footnotesize
\textit{Note:} All tests two-tailed; all $p<0.001$. Cohen's $d$ reported for accuracy comparisons; values $>$0.8 indicate large effects. CI = 95\% confidence interval for accuracy difference (percentage points). 
%% TODO: @Zhenan, compute and fill in Cohen's d for remaining rows, and add 95% CIs for accuracy differences.
\end{table}

The large effect sizes confirm that fusion improvements are practically meaningful, not merely statistically significant. However, with only five folds the $t$-tests have limited degrees of freedom ($df = 4$), and the non-independence of cross-validation folds means that $p$-values should be interpreted as approximate. The consistent direction and large magnitude of improvements across all folds and metrics, combined with the per-activity and per-subject analyses, provide converging evidence that fusion gains are robust. Subject-level cross-validation (leave-one-subject-out) is an important future direction that will require a larger participant cohort.

\section{Discussion}
\subsection{Motor Assessment Utility is Task-Dependent}
This study set out to characterize \emph{when} each sensing modality is most informative across a heterogeneous motor battery, a question that prior EEG--IMU work has left largely unaddressed~\cite{chen2021deep,grana2020improved}. The results confirm that modality utility is strongly task-dependent rather than uniform. While IMU achieved slightly higher overall accuracy (94.41\% vs.\ 92.82\%; Table~\ref{tab:overall}), activity-level analysis reveals an asymmetric pattern: IMU outperformed EEG for 7 of 10 activities, EEG showed a $3\times$ error advantage for rhythmic cycling, and one activity (seated leg extensions) reached parity (Table~\ref{tab:activity}, Fig.~\ref{fig:tradeoff}).

This asymmetry provides practical guidance for assessment design. IMU serves as a strong default for movements producing kinematically distinctive wrist-level signatures (e.g., seated side bends, wall push-ups, heel-to-toe walk), while EEG provides measurable benefit for movements where temporal pacing or rhythmic structure carries discriminative information that wrist kinematics do not capture. The key contribution is not that one modality is universally superior, but that task-dependent strengths can be quantified, offering guidance for designing sensing strategies in comprehensive motor evaluation where multiple aspects of motor function must be probed \cite{espay2016technology, del2016free, brognara2019assessing}. These complementarity patterns reflect the wrist-worn IMU configuration used here, specifically a smartwatch worn on the dominant wrist, consistent with the consumer wearable form factor most commonly deployed for real-world PD monitoring; different sensor placements would likely shift the observed modality balance.

\subsection{Neural Signatures of Rhythmic Movement}
The EEG advantage for light stationary cycling, the largest single-activity modality gap in the battery, merits mechanistic discussion. Cycling involves repetitive, temporally regular lower-limb movements at a consistent cadence with minimal trunk displacement, producing weak wrist-level acceleration signatures but potentially strong cortical oscillatory patterns.

This interpretation is supported by established neurophysiological evidence. Sensorimotor beta oscillations (13--30~Hz) exhibit characteristic event-related desynchronization during movement and post-movement rebound \cite{kilavik2013ups}, with recent work showing that beta dynamics modulate systematically during sustained physical exertion \cite{giustiniani2025modulation}. During cycling specifically, Enders et al. \cite{enders2016changes} demonstrated that alpha, beta, and gamma oscillations display alternating desynchronization and synchronization patterns time-locked to the pedaling cycle, and beta-band corticomuscular coherence has been shown to modulate during rhythmic tasks \cite{peng2024neuromechanical}. These findings suggest that rhythmic movements impose regular temporal structure on cortical oscillations that the EEGNet--Transformer architecture can exploit through its attention mechanism, which models periodic dependencies across the 4\,s window.

In contrast, cycling's weak wrist-level inertial signature renders it similar to other seated movements in IMU feature space (e.g., leg extensions: 3.88\% IMU error vs.\ cycling: 12.10\%), explaining the high IMU confusion between these activities (Table~\ref{tab:confusion}). The per-subject analysis (Table~\ref{tab:persubject}) confirms that this cycling EEG advantage is consistent across participants rather than driven by individual outliers. That beta-band abnormalities are also a hallmark of Parkinson's disease pathophysiology \cite{kilavik2013ups, giustiniani2025modulation} makes this observation clinically relevant: EEG's sensitivity to rhythmic motor patterns may be particularly valuable for detecting disrupted motor timing in movement disorder populations.

\subsection{Task-Dependent Statistical Relationships}
To further characterize modality relationships, we computed per-activity correlations between EEG and IMU window-level correctness patterns. Overall statistical agreement (Cohen's $\kappa = 0.70$) might suggest substantial overlap; however, per-activity analysis reveals task-dependent structure. Activities where both modalities perform well show high agreement (wall push-ups: $r = 0.75$), while activities where one modality struggles show low or negative correlations (seated leg extensions: $r = -0.74$; seated boxing hooks: $r = -0.71$; light cycling: $r = 0.53$).

This pattern indicates that modality independence is task-specific rather than uniform: for some movements, EEG and IMU provide largely redundant information, while for others they capture different discriminative aspects of motor behavior. The fact that aggregate $\kappa$ masks substantial per-activity variation underscores why task-level analysis, rather than battery-level summary statistics alone, is essential for evaluating multimodal sensing strategies. This finding directly supports the fusion results: integrating both modalities improves reliability precisely because the meta-learner can adaptively leverage the stronger modality for each movement context.

\subsection{Why Late Fusion Improves Reliability}
The confusion analysis (Table~\ref{tab:confusion}, Fig.~\ref{fig:confusion_diff}) reveals the mechanistic basis for fusion: dominant error modes are largely non-overlapping between modalities. EEG confuses kinematically similar seated activities (side bends $\rightarrow$ medicine ball twists), while IMU confuses activities with weak wrist-level signatures (cycling $\rightarrow$ twists/extensions). This structural dissimilarity in error patterns is a necessary condition for effective decision-level fusion \cite{baltrusaitis2019multimodal}.

Logistic regression fusion achieved the highest performance among fusion strategies (98.68\% accuracy), outperforming the MLP meta-learner (97.01\%). This suggests that the mapping from unimodal probability outputs to fused predictions is approximately linear in this setting, and that additional model capacity introduces overfitting on the limited validation-set training signal. From a clinical translation perspective, a linear meta-learner is advantageous: it is computationally lightweight, straightforward to audit, and requires minimal calibration data, properties that facilitate deployment in resource-constrained monitoring settings~\cite{baltrusaitis2019multimodal}.

The most consequential fusion benefit may be the improvement in worst-task accuracy (from $\sim$87\% to 96.76\%) and the reduction in cross-task standard deviation (from $\sim$3\% to 1.06\%). For clinical assessment batteries where consistent reliability across all measured activities is a prerequisite for valid composite scoring, this uniformity improvement ensures that no single task undermines overall assessment validity.

\subsection{Implications for Digital Motor Assessment}
The task-dependent complementarity documented here has direct implications for movement disorder assessment. PD manifests heterogeneously: some patients exhibit primarily gait-related deficits detectable through kinematic analysis \cite{espay2016technology, powers2021smartwatch, sapienza2024assessing}, while others show disrupted rhythmic motor control, notably freezing of gait, which involves breakdown of rhythmic stepping patterns \cite{sigcha2024deep, borzi2025freezing}. Our finding that rhythmic movements favor EEG while kinematically complex movements favor IMU suggests that comprehensive PD assessment batteries should incorporate multimodal sensing to capture both types of dysfunction. This aligns with emerging multimodal approaches in PD research. Bajpai et al. \cite{bajpai2023multimodal} demonstrated that EEG--IMU fusion improves freezing of gait prediction (92.1\% accuracy), and multimodal datasets integrating EEG, EMG, and accelerometry have been developed specifically for FoG detection \cite{mesin2022multimodal, abbasi2025deep}. Our contribution extends this literature by characterizing modality utility across a diverse task battery rather than a single clinical phenomenon, providing the quantitative task-level analysis needed to design optimally informative assessment protocols.

For longitudinal home-based monitoring, where wearable sensors are increasingly deployed \cite{powers2021smartwatch, sapienza2024assessing}, the subject-dependent evaluation protocol used here is directly applicable: patients would provide initial calibration data to personalize their models, after which continuous monitoring would operate in the subject-dependent regime. 

\subsection{Methodological Considerations}
Several design choices merit explicit discussion. The IMU branch uses handcrafted features with XGBoost rather than end-to-end deep learning. Feature-based pipelines remain competitive for moderate-sized wearable datasets \cite{chen2021deep, wang2022deep} and provide interpretable descriptors: the stability of selected features across folds (48/60 consistent) enables identification of which movement characteristics drive classification. Meanwhile, the EEG branch builds on EEGNet's physiologically motivated architecture \cite{lawhern2018eegnet} with transformer-based temporal modeling \cite{zhao2024ctnet, keutayeva2024compact}, an established combination for motor EEG classification.

A deliberate choice was to preserve broadband EEG without bandpass filtering or artifact rejection, allowing the model to learn task-relevant representations from normalized input. While this avoids information loss from aggressive preprocessing, it introduces an important interpretive caveat: the EEG branch may partly rely on movement-artifact-related features (e.g., electrode motion, muscle activity) rather than purely neural oscillatory activity \cite{mahmud2023mlmrs, kalita2024aneeg}. The cycling EEG advantage could reflect either genuine cortical oscillatory patterns \cite{enders2016changes, peng2024neuromechanical} or the fact that cycling produces distinctive artifact signatures at the scalp. Disambiguating these contributions through systematic frequency-band analysis or artifact-aware training \cite{mahmud2023mlmrs, kalita2024aneeg} is an important direction for future work. For the present proof-of-concept, the practically relevant finding is that broadband EEG \emph{representations}, regardless of whether their discriminative content is neural, artifactual, or both, complement IMU representations in a task-dependent manner.

\subsection{Limitations and Future Work}
This proof-of-concept study has several limitations. The cohort comprises six healthy adults, and while per-subject analyses (Table~\ref{tab:persubject}) confirm consistent task-dependent patterns, clinical validation with PD populations and larger samples is essential. The window-level cross-validation characterizes within-session discriminability but does not assess cross-session or cross-subject generalization; non-overlapping 4\,s segments and per-recording normalization mitigate but do not eliminate within-recording dependencies, and leave-one-subject-out evaluation remains an important future direction. Minor protocol deviations across participants were accounted for in labeling. Finally, the no-filtering EEG strategy means classification may partly rely on movement-artifact features rather than purely neural activity; frequency-band analysis is needed to disambiguate these contributions.

Future work will focus on validating task-dependent modality patterns in PD cohorts, evaluating cross-subject generalization via domain adaptation \cite{zhi2025supervised, li2025cross}, systematically comparing artifact-aware EEG preprocessing \cite{mahmud2023mlmrs, kalita2024aneeg} against raw-input approaches, and exploring feature-level and attention-based fusion strategies within broader multimodal healthcare frameworks \cite{krones2024review, hao2025multimodal}.

\section{Conclusion}
This study characterized task-dependent modality effectiveness across a ten-activity motor battery using synchronized EEG and wearable IMU signals. The central finding is that modality utility is activity-specific: IMU provided advantages for 7 of 10 kinematically distinctive movements, while EEG achieved $3\times$ lower error for rhythmic cycling, consistent with established evidence on cortical oscillatory engagement during repetitive movement. Decision-level fusion via logistic regression resolved these asymmetries, achieving 98.68\% accuracy, improving worst-task performance from $\sim$87\% to 96.76\%, and compressing cross-task variability to 1.06\%. Critically, the non-overlapping error structures between modalities provide a principled basis for these gains rather than mere averaging. As a proof-of-concept with six healthy participants, these results establish a methodological foundation, including per-task complementarity analysis and reliability-focused evaluation metrics. The quantitative demonstration that no single modality reliably covers all motor tasks motivates multimodal sensing strategies for movement disorder assessment. Future validation in Parkinson's disease cohorts will determine whether these task-dependent patterns translate to clinically meaningful monitoring improvements.

\section{Acknowledgments}
We acknowledge the contributions of Harsh Patel to data collection and early experimental development. We thank all participants for their time and effort during the study sessions.

\bibliographystyle{IEEEtran}
\bibliography{refs}
\end{document}